\documentclass{elsart}

\usepackage{amsmath}
\usepackage{amssymb}
\usepackage{graphicx}
\usepackage{dcolumn}
\usepackage{bm}

\def\be{\begin{eqnarray}}
\def\ee{\end{eqnarray}}
\def\ba{\begin{array}}
\def\ea{\end{array}}

\begin{document}

\begin{frontmatter}

\title{The self-consistent calculation of the edge states at quantum Hall
effect (QHE) based Mach-Zehnder interferometers (MZI)}

\author[l1]{A. Siddiki},
\author[l2]{A. E. Kavruk},
\ead{aekavruk@selcuk.edu.tr}
\author[l2]{T. \"{O}zt\"{u}rk},
\author[l2]{\"{U}. Atav},
\author[l2]{M. \c{S}ahin} and
\author[l3]{T. Hakio\u{g}lu}
\address[l1]{Physics Department, Arnold Sommerfeld Center for Theoretical Physics, and Center for NanoScience
Ludwig-Maximilians-Universit\"at M\"unchen, D-80333 Munich,
Germany}
\address[l2]{Selcuk University, Physics Dept. 42075
Konya, Turkey}
\address[l3]{Bilkent University, Physics Dept. 06690
Ankara, Turkey}

\begin{abstract}
The spatial distribution of the incompressible edge states (IES)
is obtained for a geometry which is topologically equivalent to an electronic Mach-Zehnder
interferometer, taking into account the electron-electron
interactions within a Hartree type self-consistent model. The
magnetic field dependence of these IES is investigated and it is
found that an interference pattern may be observed if two IES
merge or come very close, near the quantum point contacts. Our
calculations demonstrate that, being in a quantized Hall plateau
does not guarantee observing the interference behavior.
\end{abstract}
\begin{keyword}
edge states \sep Quantum Hall effect \sep screening \sep electronic
Mach-Zehnder interferometer
\PACS 73.20.Dx, 73.40.Hm, 73.50.-h, 73.61,-r
\end{keyword}
\end{frontmatter}
%

The puzzling interference patterns observed at the QHE based
electronic MZI~\cite{Heiblum03:415} setups have already attracted
many theoreticians to investigate the structure of the "edge
states" at these samples. The realistic modelling of the
electrostatic potential and electronic density distributions is
believed to be indispensable in understanding the rearrangement of
the edge states involved. Therefore, the electron-electron
interaction has been proposed~\cite{Florian05:788,Neder07:112} as
a possible source of dephasing in these experiments. It was stated
that, the conventional edge state explanation of the QHE, i.e.
Landauer-B\"uttiker (LB) formalism, fails to cover the
experimental findings. On the other hand, a detailed analysis of
the QHE related physics, taking account the formation of the
incompressible strips, is needed for a direct comparison with
experimental data. Recently, a two channel edge state model is
proposed~\cite{Neder07:112}, which is able to explain the observed
visibility oscillations in terms of a non-Gaussian noise. The
essential parameters are the electron velocity and the coupling
strength between the "interference" and "detector" channels. In
this paper, we extend a previous work~\cite{SiddikiMarquardt} to
investigate the electrostatics of e-MZI setups in the integer QHE,
assuming a topologically equivalent geometry to the experimental
one~\cite{Samuelson04:02605}. We aim to provide explicit
calculations of the spatial rearrangement of the incompressible
edge states~\cite{siddiki2004}. The widely used self-consistent
Thomas-Fermi-Poisson screening theory~\cite{SiddikiMarquardt} is
used to obtain the electron density and electrostatic potential.
We propose two possible scenarios to observe interference,
depending on the distribution of the incompressible strips, in
other words depending on the magnetic field strength.

To obtain the confinement potential, we follow the procedure
proposed by Ref.~\cite{Davies94:4800}. In this model the bare
potential can be obtained at the level of two dimensional electron
system (2DES), i.e. in the plane of $z=z_0$ measured from the
surface into the sample, provided that the surface gate pattern
and the potential distribution are known. The contribution of the
gates to the total potential at the 2DES is given by \be
\label{eq:davies}V_{\rm{gate}}({\bf{r}},z_{0})=\frac{1}{\kappa}
\int{\frac{|z_{0}|}{2\pi(z_{0}^{2}+|{\bf{r}}-{\bf{r'}}|^{2})^{3/2}}}V_{\rm
g}({\bf{r'}},0)d{\bf{r'}} \ee where $V_{\rm g}({\bf{r'}},0)$ is
the potential on the sample surface. The second contribution to
the external potential comes from the donors, which we simulate by
a half-period cosine function in the $y$ direction. Given the
external potential in the plane of 2DES; $V_{ext}({\bf
r},z_0)=V_{gate}({\bf r},z_0)+V_{donor}({\bf r},z_0)$,
$(x,y,z_0)=({\bf r},z_0)$, in the real space, it is
straightforward to calculate the screened potential (again in the
real space) at $T=0$ and $B=0$, making two dimensional forward and
back Fourier transform, using $V_{scr}(q)=V_{ext}(q)/\epsilon(q)$,
where $\epsilon(q)$ is the momentum $(q)$ dependent Thomas-Fermi
dielectric function defined by $\epsilon(q)=1+2/a_{B}^{*}|q|$, and
$a_{B}^{*}$ is the effective Bohr radius ($\sim 10$nm). We use
this screened potential as an initial input for the following set
of two self-consistent (SC) equations: \be \label{thomas-fermi}
 n_{\rm el}({\bf{r}})=\int dE\,D(E)f\big( [E+V({\bf{r}})-\mu^{\star}]/k_{B}T \big),
\ee
for the spinless electron density, where $D(E)$ is the bare Landau
density of states (DOS), $f(\alpha)=[1+e^{\alpha}]^{-1}$ the Fermi
function and the Hartree potential energy of an electron \be
\label{hartree} V_{H}({\bf{r}})= \frac{2e^2}{\bar{\kappa}}
\int_{A} \!d{\bf{r'}} K({\bf{r}},{\bf{r'}})\,n_{\rm
el}({\bf{r'}}), \ee within the Thomas-Fermi approximation. In the
case of periodic boundary conditions, as we consider, the kernel
$K({\bf{r}},{\bf{r'}})$ can be expressed in an analytically closed
form~\cite{Morse-Feshbach53:1240}, otherwise has to be obtained
numerically. The total potential energy is obviously nothing but
the sum of Hartree and external potential energies.
\begin{figure}
{\centering
\includegraphics[width=.7\linewidth]{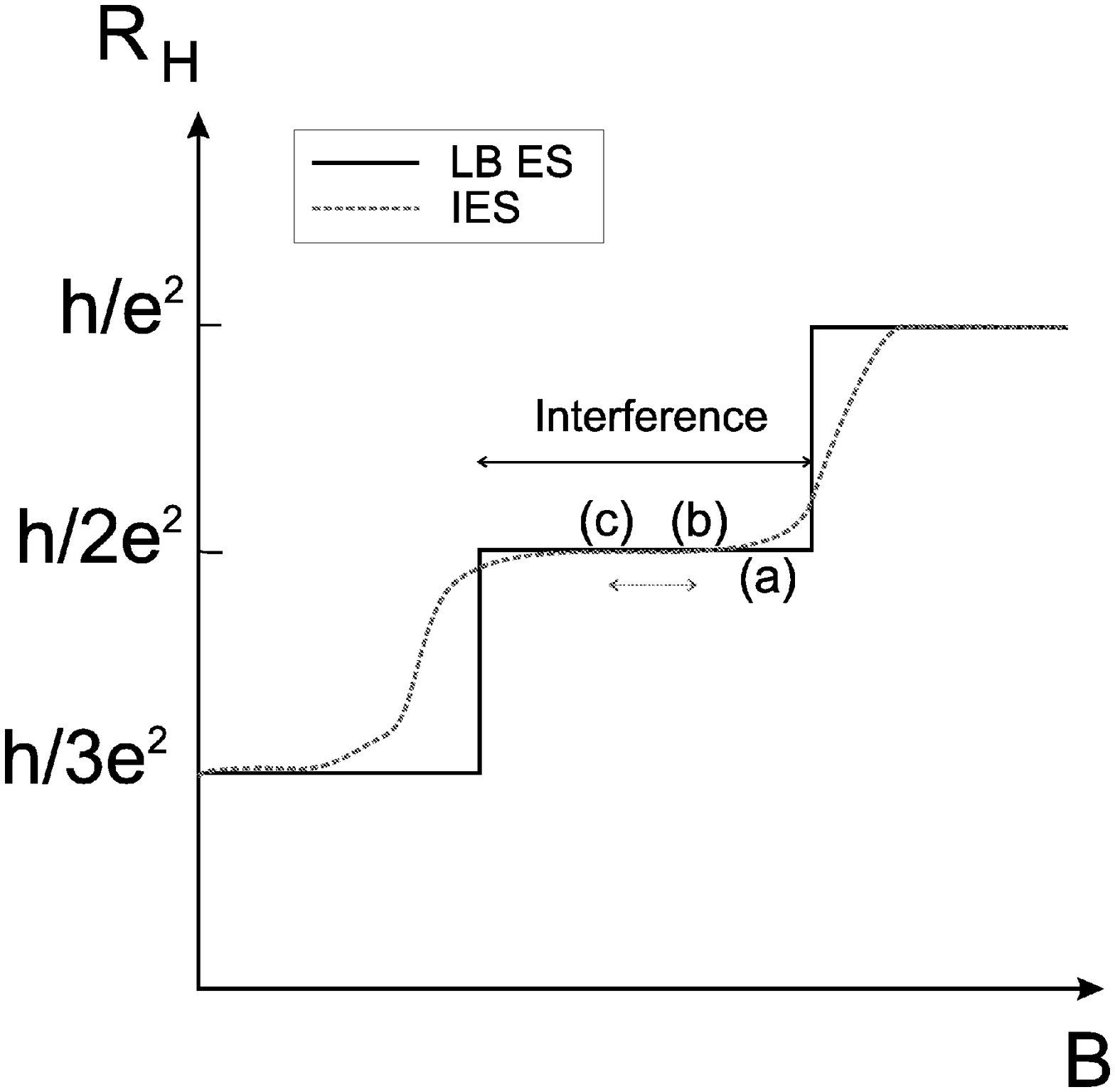}
%
\caption{ \label{fig:fig1} The sketch of the Hall resistance,
considering LB ES (solid line) and incompressible ES (broken
line). Expected interference intervals of the magnetic field
are denoted by the lines with arrows on both ends.}}
\end{figure}
In the conventional edge state explanation of the integer QHE, one
counts the number of the LB ES, which essentially gives the
plateau number with an integer filling factor. Without assuming
any sort of localization or disorder, the Hall resistance looks
like a staircase (cf. Fig. \ref{fig:fig1}), whereas longitudinal
resistance exhibits delta spikes at transitions as a function of
the $B$ field. This implies that, whenever one enters to a plateau
region one always have a percolating LB ES from source to drain,
therefore the interference pattern should be observed throughout
all the plateau region, contradicting with the experimental
findings. Although such a discrepancy can be removed by a large
amount of (asymmetric) Landau level broadening, the high mobility
of the sample rules out this possibility. On the other hand, in
the screening theory of the QHE~\cite{siddiki2004,Siddiki:ijmp} a
plateau occurs only if at least an IES exists along the current
direction. Within this localization free model, the widths of the
plateaus are limited by the thicknesses of the IES, depending on
the $B$ field and/or the long-range part of the
disorder~\cite{Siddiki:ijmp}. In Fig.~\ref{fig:fig2}, we plot the
electron distribution as a function of the spatial coordinates,
calculated within the above SC scheme for a typical Fermi energy,
$E_F=12.75$ meV at $1^{o}$ K. \begin{figure} {\centering
\includegraphics[width=1.0\linewidth]{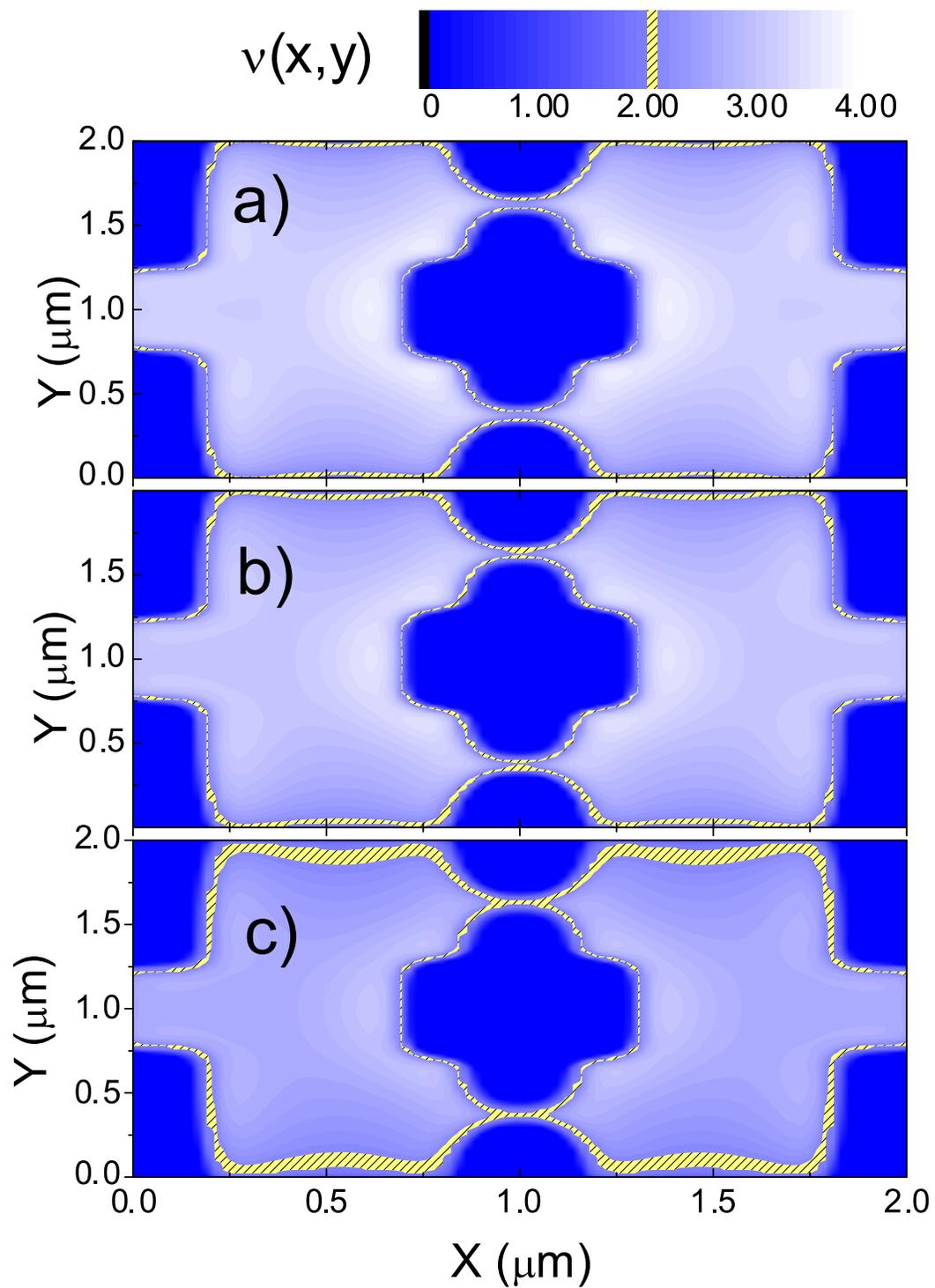}
%
\caption{ \label{fig:fig2} The color coded $\nu(x,y)$ for
$\Omega=1.15,0.95,0.90$. The calculations are done at
$\Omega/k_{B}T \approx 0.025$. The gates defining the geometry,
are taken to be $85$ nm above the 2DES and biased with $-1.0$ V.
}}
\end{figure}The color gradient depicts from zero
(dark) to high electron concentration. The high potential bias at
the surface guarantees that no electrons can reside under the
gates, whereas the light (yellow) stripes highlight the positions
of local filling factor two. Under the conditions considered at
the top most panel of Fig. \ref{fig:fig2} the system will be
observed to be almost entering to the plateau region since there
exists two (almost) percolating IES (at the top and the bottom
part of the geometry), however, we believe that the visibility
will be either too small to observe (due to scattering at the
constriction or at the bulk) or will be zero. The middle panel
shows a situation that, the percolation of the IES is well formed
and the system is on the plateau, on contrast the visibility will
still be small, since the interference will continue to be
dominated by the tunnelling where scattering processes may take
place. For the lowest $B$ field, the IES merge at the quantum
constrictions and decoherence is minimized, therefore the
visibility is predicted to be the highest at a reasonably high
mobility sample, similar to the samples measured at the
experiments~\cite{Heiblum03:415,Neder07:112}. We believe that, if
(not only if) the non-dissipative current is confined to the IES,
where no backscattering takes place, the observed amplitude
variation of the visibility as a function of $B$ field at the
experiments can simply be explained by an emerging IES at the
QPCs. This claim also promotes the fact that, in such sensitive
experiments the geometrical shape of the QPCs may play an
important role~\cite{SiddikiMarquardt,engin07}, although the
transmission amplitude remains unchanged.

In summary, the spatial distribution of the back-scattering free
IES at a MZI (topologically equivalent geometry) is studied,
exploiting the smooth variation of the external potential, within
the Thomas-Fermi approximation. We have shown that, one would not
observe visibility oscillations on each and single magnetic value
of the quantized Hall plateau interval in contrast to the LB ES
model. We have reasoned this on the base of merging IES at the
QPCs and highlighting the importance of the geometry of the
constriction. The amplitude itself and the interference interval
clearly depend on the mobility of the sample, therefore (also
including spin) an extension of this present work may help in
improving the sample design and the quality of the observed
quantities.

The authors acknowledge the support of the Marmaris Institute of
Theoretical and Applied Physics (ITAP), TUBITAK grant 105T110, and
Selcuk University BAP contracts 07101003 \& 07101037, SFB631 and
DIP.

\end{document}